\documentclass[twocolumn]{aa}

\usepackage{graphicx}
\usepackage[varg]{txfonts}
\usepackage[colorlinks]{hyperref}
\usepackage{natbib}
\bibpunct{(}{)}{;}{a}{}{,} 
\usepackage{lscape}
\usepackage[normalem]{ulem}
\usepackage{color}

\begin{document}

\title{Comparing extrapolations of the coronal magnetic field structure at 2.5~$R_\sun$ with multi--viewpoint coronagraphic observations}

\author{C. Sasso\inst{1} \and R. F. Pinto\inst{2} \and V. Andretta\inst{1} \and R. A. Howard\inst{3} \and A. Vourlidas\inst{4} \and A. Bemporad\inst{5} \and S. Dolei\inst{6} \and D. Spadaro\inst{6} \and R. Susino\inst{5} \and E. Antonucci\inst{5} \and L. Abbo\inst{5} \and V. Da Deppo\inst{7} \and S. Fineschi\inst{5} \and F. Frassetto\inst{7} \and F. Landini\inst{8} \and G. Naletto\inst{9,7} \and G. Nicolini\inst{5} \and P. Nicolosi\inst{9} \and M. Pancrazzi\inst{8} \and M. Romoli\inst{10} \and D. Telloni\inst{5} \and R. Ventura\inst{6}}

\offprints{C. Sasso, \email{csasso@oacn.inaf.it}}

\institute{INAF--Osservatorio Astronomico di Capodimonte, Salita Moiariello 16, I--80131 Napoli, Italy \and Institut de Recherche en Astrophysique et Planétologie, University of Toulouse, CNRS, UPS, CNES, 9 avenue Colonel Roche, BP 44346-31028, Toulouse Cedex 4A, France \and Space Science Division, U.S. Naval Research Laboratory, Washington, DC, USA \and Johns Hopkins University Applied Physics Laboratory, Laurel, MD, USA \and INAF--Turin Astrophysical Observatory, Via Osservatorio 20, 10025 Pino Torinese (TO), Italy \and INAF--Catania Astrophysical Observatory, Via Santa Sofia 78, 95123 Catania, Italy \and CNR--Institute of Photonics and Nanotechnologies, via Trasea 7, 35131 Padova, Italy \and INAF--Arcetri Astrophysical Observatory, Largo Enrico Fermi 5, 50125 Firenze, Italy \and University of Padova, Department of Physics and Astronomy "Galileo Galilei"
Via Marzolo, 8 I-35131 Padova, Italy \and University of Florence--Dept. of Physics and Astronomy, Largo Enrico Fermi 2, 50125 Firenze, Italy} 

\date{Received / Accepted}

\abstract{The magnetic field shapes the structure of the solar corona but we still know little about the interrelationships between the coronal magnetic field configurations and the resulting quasi--stationary structures observed in coronagraphic images (as streamers, plumes, coronal holes). One way to obtain information on the large--scale structure of the coronal magnetic field is to extrapolate it from photospheric data and compare the results with coronagraphic images. Our aim is to verify if this comparison can be a fast method to check systematically the reliability of the many methods available to reconstruct the coronal magnetic field. Coronal fields are usually extrapolated from photospheric measurements typically in a region close to the central meridian on the solar disk and then compared with coronagraphic images at the limbs, acquired at least 7 days before or after to account for solar rotation, implicitly assuming that no significant changes occurred in the corona during that period. In this work, we combine images from three coronagraphs (SOHO/LASCO--C2 and the two STEREO/SECCHI--COR1) observing the Sun from different viewing angles to build Carrington maps covering the entire corona to reduce the effect of temporal evolution to $\sim 5$ days. We then compare the position of the observed streamers in these Carrington maps with that of the neutral lines obtained from four different magnetic field extrapolations, to evaluate the performances of the latter in the solar corona. Our results show that the location of coronal streamers can provide important indications to discriminate between different magnetic field extrapolations.}

\keywords{}

\titlerunning{Comparing extrapolations of coronal field with multi--viewpoint observations}

\maketitle

\section{Introduction}\label{intro}

It is well--known that the magnetic field of the Sun drives the dynamics and structure of the solar corona. Eclipses and coronagraphic images reveal that the plasma in the corona is organized in long--lived structures, such as streamers, coronal holes, plumes, etc., which follow the configuration of the large scale magnetic field. However, the details of the interplay between plasma and magnetic field are often hard to establish. One way to obtain information on the large--scale structure of the coronal magnetic field is to extrapolate it from photospheric data and compare the results with coronagraphic images. Usually these extrapolations are based on photospheric field measurements that are acquired in a region close to the central meridian on the solar disk and then compared with coronal structures observed above the limbs. Nevertheless, this comparison requires to assume that no significant changes occurred in the global distribution of large scale features in the solar corona over about 7 days, to account for one quarter of solar rotation.

The purpose of this work is to show how coronagraphic white--light images could provide (or not) additional boundary conditions for the extrapolated fields. This kind of check is particularly important for space missions for which the connectivity problem is crucial. Quick but reliable checks like the one that will be discussed in this paper will help observation planning, even with relatively little data (i.e., "low latency data" or "beacon") and short computational time. The work described here was, indeed, part of the activities performed for the "Modeling and Data Analysis Working Group (MADAWG)", that is aimed at optimizing the coronal magnetic field extrapolations to establish the magnetic connectivity of the Solar Orbiter spacecraft \citep{muller} with the Sun, to relate future remote sensing and in situ observations. Alternative and more sophisticated methods such as tomography would of course provide a more complete view of the distribution of the white--light features. We need, however, to minimize the computational time and the amount of data to be analyzed routinely (on a daily basis). 

It has been long recognized that the streamers seen in white light coronagraphic images represent edgewise views of the coronal plasma surrounding the coronal current sheet, which is rotating with the Sun \citep[e.g.,][]{howard,bruno,wilcox}. This position of the current sheet can be estimated by a potential field calculation for a source surface at 2.5~$R_\sun$ \citep{hoeksema}. The source surface is defined as the region where currents in the corona cancel the transverse magnetic field \citep{schatten}. \citet{koomen} used images of the corona beyond 2.5~$R_\sun$, from March 1979 (before solar maximum) to September 1985 (beginning of solar minimum) and potential field extrapolations to confirm this idea. In that case, the computed magnetic neutral line at a source surface of 2.5~$R_\sun$ was defining a relatively flat current sheet at the minimum period, when the solar magnetic field was dipole-like but also near solar maximum when they report the presence of a current sheet but strongly tilted to the heliographic equator. 

\citet{wang5} compared the Large Angle and Spectrometric COronagraph \citep[LASCO;][]{brueckner} on--board the SOlar and Heliospheric Observatory \citep[SOHO;][]{domingo} white--light Carrington maps to Potential Field Source Surface (PFSS) extrapolations during 1996 solar minimum activity phase. They found that the topological appearance and evolution of the coronal streamer belt can be described as the line--of--sight viewing of a warped plasma sheet encircling the Sun and not as localized enhancements of the coronal density. At larger heliospheric distances this current sheet is observed in situ as the heliospheric current sheet (HCS). \citet{wang7} repeated the analysis for observations close to solar maximum providing further support for the idea that the coronal streamer belt beyond 2.5~$R_\sun$ is a narrow plasma sheet seen in projection outlining the HCS. With the emergence of strong, non-axisymmetric fields in the sunspot latitudes during 1998, the HCS became progressively more tilted and warped. \citet{liewer} addressed the same question of whether the streamers are the results of scattering from regions of enhanced density or the result of line--of--sight viewing of a warped current sheet. They analyzed 1.5 months of coronagraphic observations to determine the 3--D location of several bright stable streamers in the outer corona and their relationship to the coronal magnetic field through potential field extrapolations from photospheric field measurements. The comparison of the streamers' locations with that of the current sheet showed that all of the streamers lie on or near the heliospheric current sheet. To explain the presence of discrepancies between synthetic maps and observations they proposed that additional fine streamers result from flux tubes containing plasma of higher density and not from folds in the plasma sheet. 

Working on the same observations as \citet{wang5}, \citet{zhao} compared the magnetic neutral line computed from coronal magnetic field extrapolations with the position of the coronal streamer belt, finding a good match at various heights. Also \citet{saez1} investigated the 3--D structure of the solar corona comparing synoptic maps of the streamer belt obtained with the LASCO-C2 coronagraph and the simulated synoptic maps constructed from a model of the warped plasma sheet. The position of the neutral line at the source surface $(2.5~R_\sun)$ was determined using a potential field source surface model. Through this analysis they generally confirmed the earlier findings of \citet{wang5} that the streamers are associated with folds in the plasma sheet. For the fine features visible in the LASCO synoptic maps that cannot be reproduced with a model like the one of \citet{wang7}, they propose that two types of large--scale structures take part in the formation of these additional features. The first one is an additional fold of the neutral line, which does not appear in the modelled source surface neutral line, but is well visible in photospheric magnetograms. The second one is a plasma sheet with a ramification in the form of a secondary short plasma sheet. 

More recently, \citet{wang6} identified a new streamer-like structure in the corona, the so-called pseudostreamers, that separate coronal holes of the same polarity, overlying twin loop arcades without a current sheet in the outer corona, while helmet streamer overlie a single (or an odd number of) loop arcades in the lower corona and they have oppositely oriented open magnetic field in the upper corona, such that a current sheet is present between the two open field domains \citep[see Fig.~1 of][]{rachmeler}. In other words, pseudostreamers do not represent folds of the large scale coronal magnetic field. 

One major limitation of these studies is their reliance on synoptic magnetic maps accumulated over a solar rotation. Obviously, the assumption that the corona remains unchanged over 27 days becomes weaker away from solar cycle minimum leading to the discrepancies with the white light observations we discussed earlier. With the operation of the two Sun-Earth Connection Coronal and Heliospheric Investigation \citep[SECCHI;][]{secchi} COR1 coronagraphs aboard the twin Solar TErrestrial RElations Observatory \citep[STEREO;][]{stereo} spacecraft, Ahead and Behind in 2007, and the continuing LASCO operations, it has become possible to obtain an almost instantaneous picture of the corona with the minimum amount of temporal evolution, by combining the coronagraphic images from different viewing angles. These maps could then be used to evaluate the results of magnetic field extrapolations. 

As mentioned earlier, similar studies done in the past comparing extrapolations with coronal features had normally to face the additional uncertainties introduced by the need of using synoptic coronal maps built over an entire solar rotation. In this work, we aim to reduce at a minimum the uncertainties on the observational side of the comparison by exploiting a particularly favourable configuration of the SOHO and the two STEREO spacecraft. Therefore, we combine images from the COR1s and LASCO/C2 instruments for the Carrington Rotation (CR) 2091 (2009 December 07 -- 2010 January 03) and compare the Carrington map obtained with magnetic field extrapolations. In the following, we use the names of the three spacecraft SOHO, STEREO-A (STA) and STEREO-B (STB) to refer to the respective coronagraphic observations.  

The paper is organized as follows: in Sec.~\ref{sec:obs} we describe the data we used and the method adopted to merge multi-spacecraft Carrington maps into near-synchronic maps of the corona at a given date; in Sec.~\ref{sec:extrapolations} we describe the calculations of the position of the neutral line at the source surface; in Sec.~\ref{sec:results} we compare the observations and the calculations and discuss the main results, drawing our conclusions in Sec.~\ref{sec:conclusions}.

\section{Observations and Data Analysis}\label{sec:obs}

To make our analysis more relevant to the Solar Orbiter mission, currently scheduled for launch in 2020 by the European Space Agency (ESA), we test our technique on a coronal configuration similar to the coronal structure expected for the early phase of the Solar Orbiter mission. In particular, we chose CR 2091 (2009 December 07 -- 2010 January 03) as a representative time frame of the rising phase of solar cycle 24. The selected time interval had also the advantage to occur around the minimum of solar activity cycle, and no major solar eruptions (that could possibly modify the large scale coronal configuration) occurred during this period. In addition, during this Carrington rotation an active region (NOAA 11039) emerged on 27 December 2009 around Carrington longitude $\sim 90^{\circ}$, thus allowing us to test the assumption commonly made in this kind of studies that the global configuration of the corona varies little in the time frame of the order of a solar rotation.

\begin{figure}
  \centering
  \includegraphics[clip=true,angle=-90,width=8cm]{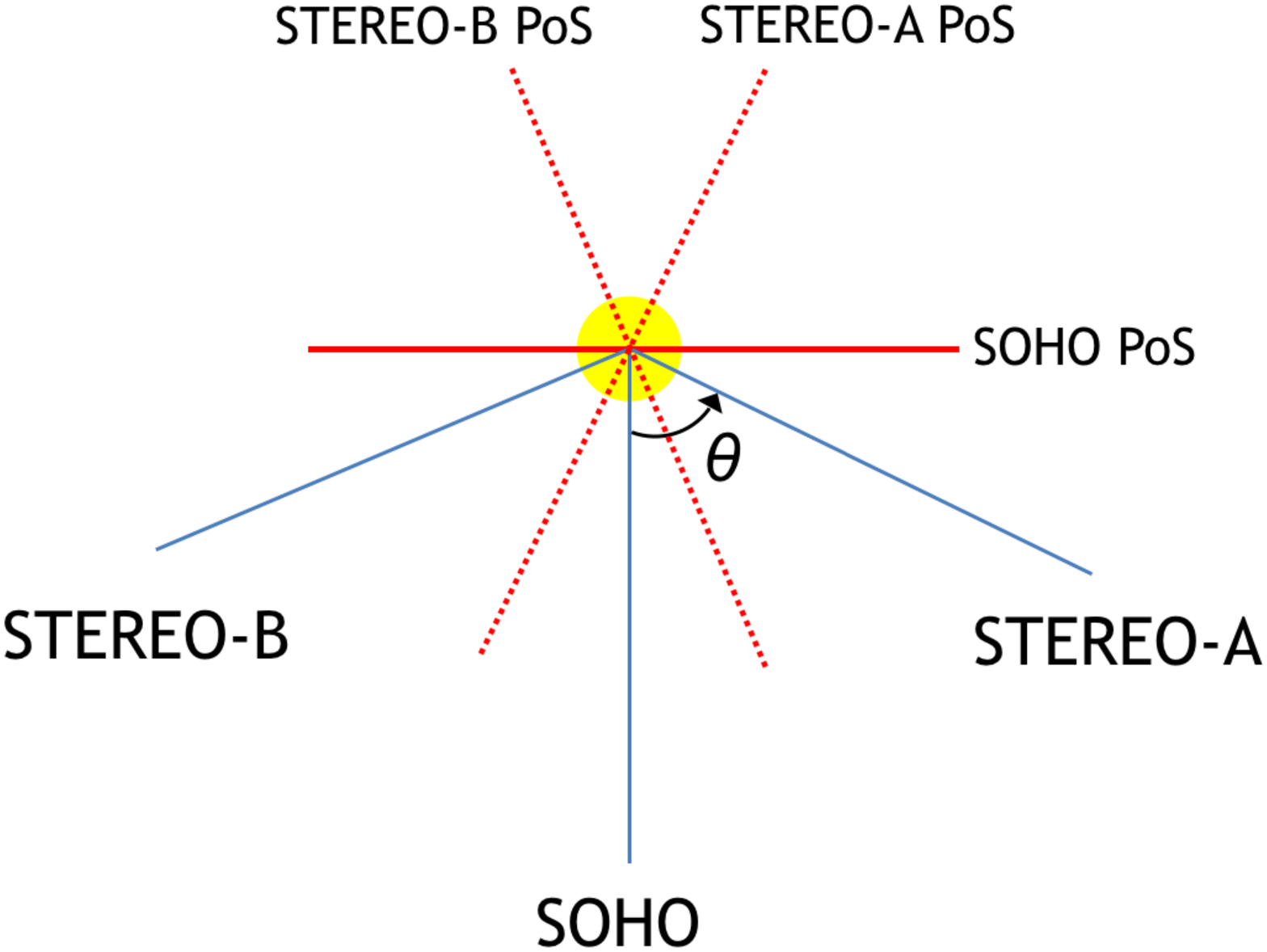}
  \caption{Relative positions of the spacecraft SOHO, STA and STB and their planes of sky (PoS, in red) with the Sun at the centre of the graph. $\theta$ is the separation angle between the SOHO and STEREO spacecraft.}
  \label{fig:coord}
\end{figure}

In Fig.~\ref{fig:coord} we show the relative positions of the three spacecraft and their planes of sky (PoS, red lines) for CR 2091, with the Sun at the centre of the graph. The STEREO--SOHO separation angles ($\theta$) were $64^\circ$ (STA) and $-67^\circ$ (STB) on 20 December 2009, 20:20 UT. The relative positions of the three spacecraft were therefore particularly favourable to scan the full corona in a time shorter than a full rotation: after $\sim 5$ days the same structures seen by SOHO are observed by STA and were observed $\sim 5$ days before by STB. 

The extrapolations are compared with the observations (see Sect.~\ref{sec:results}) at the source surface, commonly placed at a height of $r_{SS}=2.5$~R$_\sun$. At this source--surface radius, the geometry of the extrapolated magnetic field best matches the shapes of the coronal structures observed in white--light during solar eclipses, especially the size of the streamers and coronal hole boundaries \citep{wang1,wang2}. Some authors \citep[][for example]{lee} have shown that using smaller source surface heights gives improved agreement between the EUV images and the modelled open field regions during low solar activity periods. They also suggest that the source surface height is changing over time and that the energy balance may be different from one solar minimum to the next, depending on both the polar and overall photospheric field strengths as well as the open field topology. We nevertheless decided to perform the PFSS extrapolations at $2.5$~R$_\sun$ to better compare our work with the majority of past efforts. 

\begin{figure*}
  \centering
  \includegraphics[clip=true,angle=180,width=\textwidth]{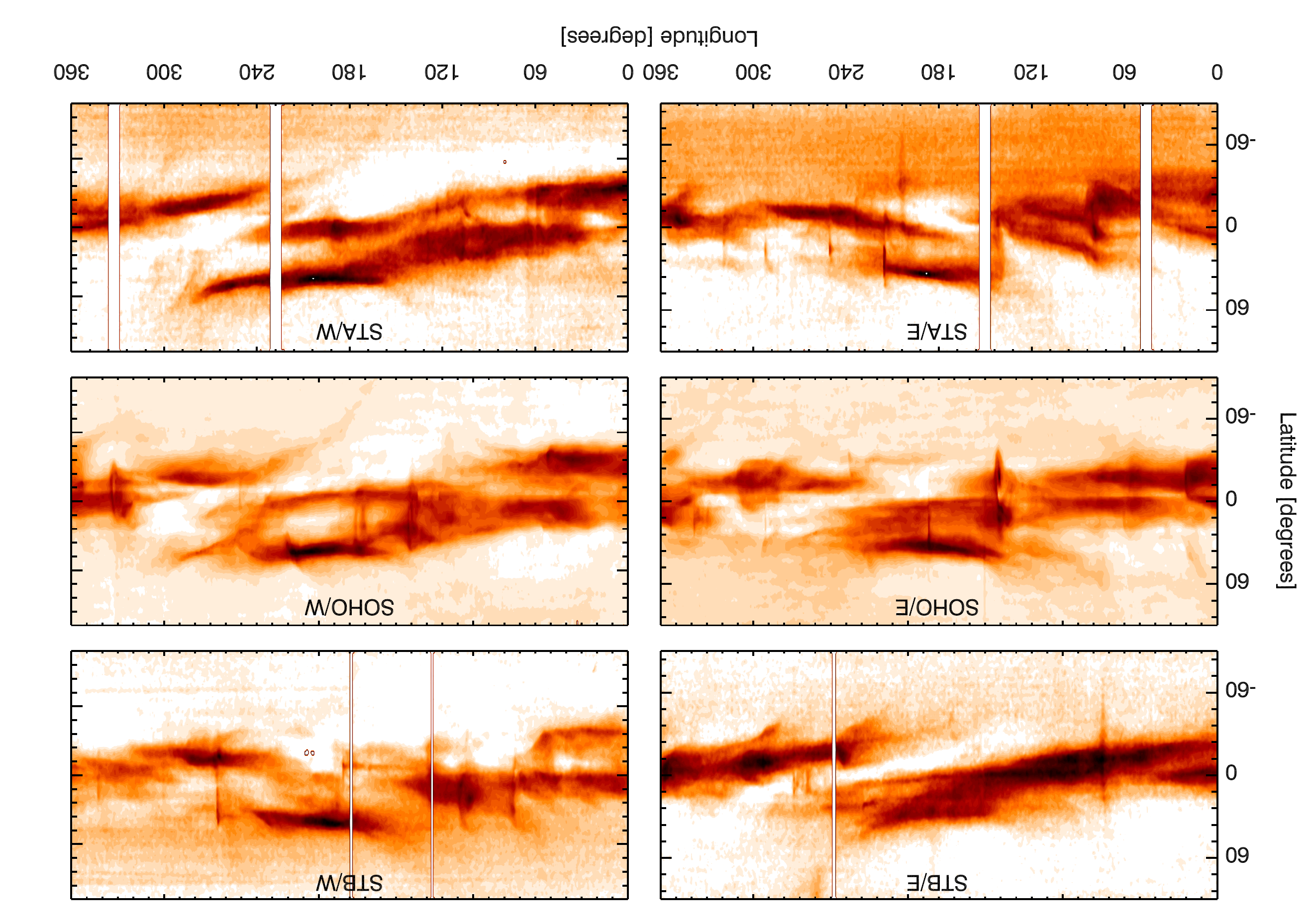}
  \caption{Carrington maps at 2.5~$R_\sun$ for CR 2091 from STB/COR1 (top), SOHO/C2 (middle), and STB/COR1 (bottom), for the East and West limb of the Sun (left and right column, respectively). Images are displayed in a reverse colour scale, i.e.\ brighter coronal features are shown in darker colours.}
  \label{fig:obs}
\end{figure*}

In Fig.~\ref{fig:obs} we show the CR 2091 Carrington maps at 2.5~$R_\sun$ from STB/COR1, SOHO/C2, and STA/COR1 data (from top to bottom) for the East and the West limb of the Sun (left and right column, respectively). In this work we represent images in a reverse colour scale, so brighter coronal features, corresponding to regions of enhanced electron density, are displayed in darker colours. We combine these maps in a single, near-synchronic synoptic map of the corona as described in the following section, thus facilitating a comparison with the results of the photospheric extrapolations described in Sec.~\ref{sec:extrapolations}. 

\subsection{Coronagraphic CR maps}\label{sec:obs:corcrmaps}

Coronagraphic CR maps result from synoptic maps built by extracting from each image a circular profile at a given heliocentric height. These annular slices are then piled up, each column representing one circular profile in the original image. In the synoptic maps, the X axis gives the time of the observations. The differences with the CR maps lie in the parametrization of the axis: the X axis gives the Carrington longitude and the Y axis the Carrington latitude. Thus, we have a map of the solar corona at a certain radial distance from the Sun centre. 

\subsection{Combined CR maps}\label{sec:obs:merging}

To combine the six CR maps from the three spacecraft, we first normalize each image to its maximum value; in the case of STA and STB images, we find necessary to subtract a constant value prior the normalization process. This ``background'' value was estimated at the 10$^\mathrm{th}$ percentile of the image histogram. Even so, a noticeable asymmetry between the two poles remains in the CR images from both STEREO spacecraft (see, for example, the top right panel in Fig.~\ref{fig:obs}, where the intensities between the positive and negative latitudes are clearly different). Inspection of some of the original coronal images, reveals that the asymmetry likely points to asymmetry of the instrumental straylight. In any case, since we are interested in the brightest features of these CR maps, we do not attempt to further correct for this effect.

\begin{figure}
\centering
\includegraphics[clip=true,width=\linewidth]{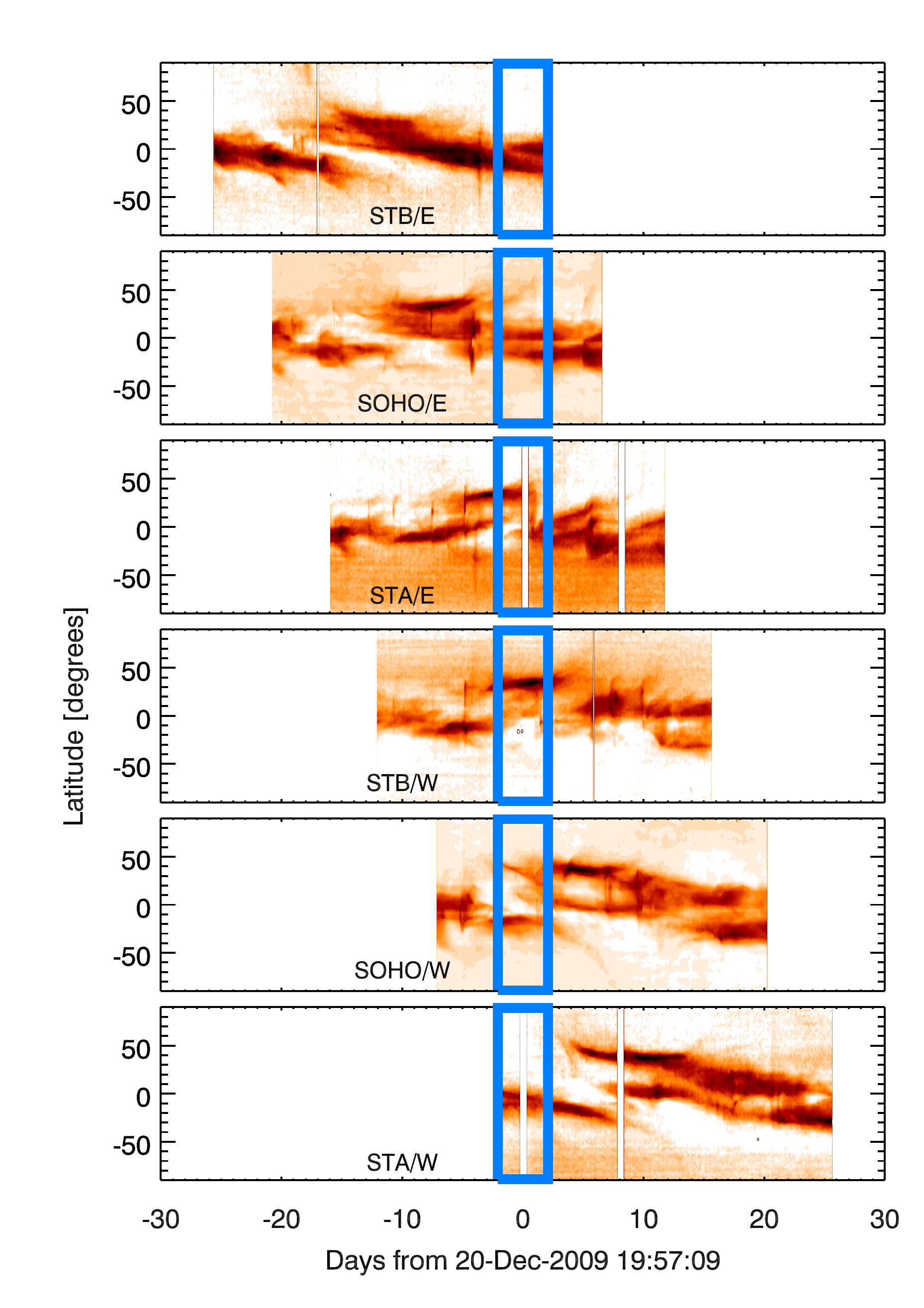}
\caption{Normalized Carrington maps (same as Fig.~\ref{fig:obs}) at 2.5~$R_\sun$, for CR 2091, from STB, SOHO, and STA, aligned to a common reference time. The blue box indicates the parts of the maps observed within $\pm$2 days from the centre date of CR 2091. By collating these six boxes (starting from STB East and ending with STA West) we can obtain a near-synchronic (to within $\sim 5$ days) map of the entire corona.}
\label{fig:obs_time}
\end{figure}

We then re-sample all images to a common longitude and latitude grid. Since each Carrington longitude in a CR map corresponds to a given observing time of the corona, the six CR maps can be displayed on a common time line, as shown in Fig.~\ref{fig:obs_time}, by means of the temporal distance, $\Delta t$, of each longitude slice with respect to a reference time. We chose 20 December 2009 20:00 UT (the centre date of CR 2091) as the reference time. Using this representation, it is easy to verify that the entire corona is observed by the three instruments over a time range of little more than 4 days around the reference date of 20 December 2009. In the blue boxes we highlight the coronal sections observed by each instrument. 

We want to determine a CR map representing the configuration of the corona in as short time interval as possible (a "near-synchronic" CR map), thus minimizing the effect of the evolution of coronal structures. With the three spacecraft in the favourable configuration shown in Fig.~\ref{fig:coord}, it is indeed possible to scan the entire corona in about 1/6$^\text{th}$ of a Carrington rotation (considering an average angle of $66^\circ$ between the spacecraft and an average rotation time of 27.3 days, the time needed to span that angle is $27.3\times66/360=5.00$ days) in an already significant improvement over the assumption underlying the typical usage of CR maps from a single vantage point, i.e. that the corona does not significantly change during a solar rotation.  

In this context, it is also useful to note that a polarized brightness (pB) observation at 2.5~$R_\sun$ of an axisymmetric coronal structure at the solar equator is the result of the integration along of the line of sight of a kernel with a full width at half maximum $\text{FWHM} \sim 45^\circ$. This angular extent is spanned by the plane of the sky (PoS) rotating with the Sun in about 5 days. We can consider this value as an upper limit of the intrinsic ambiguity in longitude of a CR map in the equatorial regions. More realistic values have been determined by, e.g., \citet{thernisien}, who presented a three-dimensional reconstruction of the electron density of a streamer, characterizing also its length and thickness. For a streamer observed by LASCO on January 2004 during CR 2012, they found that the FWHM of the streamer normalized brightness along the line of sight was $\sim 8^\circ$ at 2.5~$R_\sun$, a value corresponding to $\sim 0.6$ days. More importantly, they also found that observed changes in the streamer appearance were due to changes in the viewing geometry, while the intrinsic properties of the streamer did not significantly change over the time interval (about 7 days) needed to observe the streamer from edge-on to face-on.

With these considerations in mind, and with the obvious exception of events like Coronal Mass Ejections (CMEs), we can reasonably expect that a CR map built over $\sim 5$ days, as it is the case of our present study, is as close to a snapshot of the corona around a given date as it could be obtained with this approach. 

\begin{figure}
\centering
\includegraphics[clip=true,angle=180,width=\linewidth]{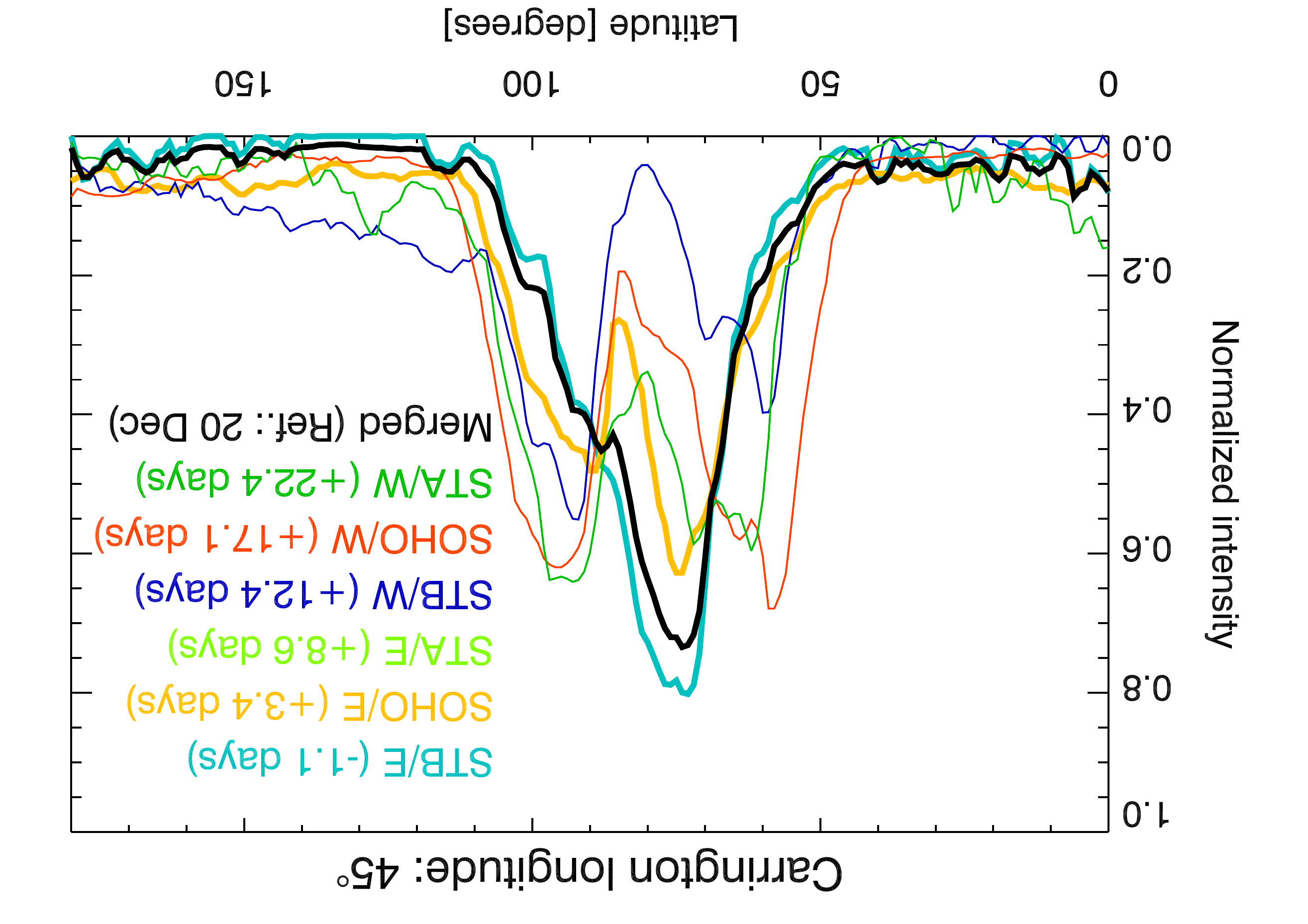}
\caption{Normalized intensity of the slices at 45$\degr$ longitude of the six Carrington maps represented in Fig.\ref{fig:obs} at 2.5~$R_\sun$ for CR 2091 (colour coding given in the figure legend), together with the weighted averaged intensity computed as described in Sec.~\ref{sec:obs:merging}, shown as a solid, black line and indicated as "Merged" in the legend. The slices closest in time to the reference time ($\Delta t_i$ at 45$\degr$ longitude, for each map, is indicated in the legend) are also shown using thicker lines. The slice contributing mostly to the merged slice (black line) is the one with $\Delta t_i$ at that longitude of 1.1 days (slice of STB/E, cyan line).}
\label{fig:comb_ex}
\end{figure}

\begin{figure}
\centering
\includegraphics[clip=true,angle=180,width=\linewidth]{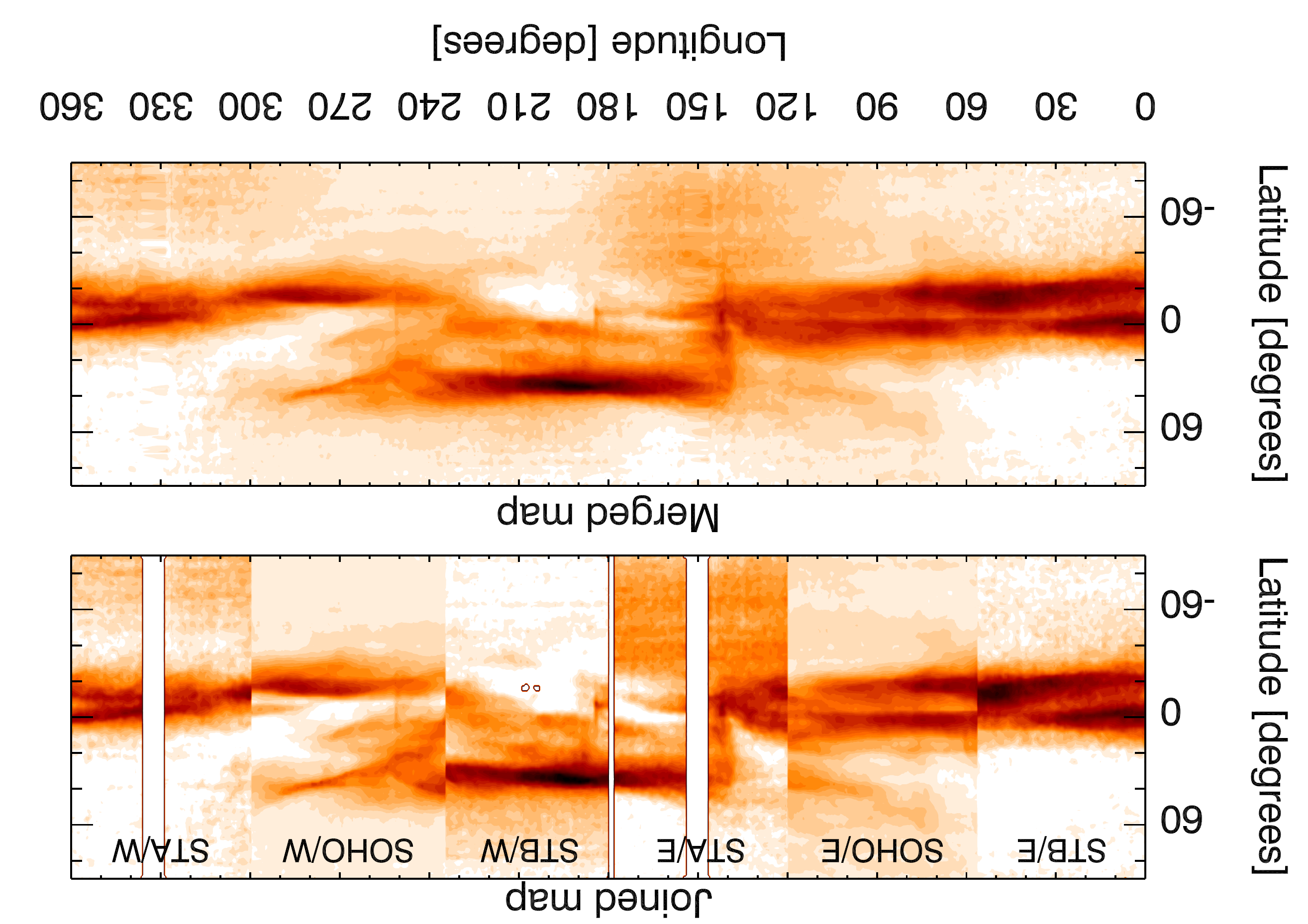}
\caption{CR 2091 maps, at 2.5~$R_\sun$ obtained by combining both West and East limb Carrington maps for all the three spacecraft with the two methods described in Sec.~\ref{sec:obs:merging}.}
\label{fig:comb3}
\end{figure}

Once the six Carrington maps were scaled to common intensity and coordinate ranges, we used the following two methods to obtain a combined Carrington map representing a near-synchronic map of the solar corona around a chosen reference time, in this case the middle time of the time interval considered (CR 2091):
\begin{enumerate}
\item (``Joined map'') At each Carrington longitude, the slice of a normalized CR map whose observing time is closest to the reference time is selected. The advantage of this method is its simplicity; the resulting combined map, however, exhibits noticeable discontinuities at the times where two spacecraft observe the same longitude at similar temporal distance from the reference time.
\item (``Merged map'') At each Carrington longitude, the weighted average of all the normalized CR maps for that longitude is computed, where the weight assigned to the $i$-th map is a function of the temporal distance to the reference time of the observation of that slice at that longitude, $\Delta t_i$. In particular, we chose weights proportional to $1/\left[1+\left(16\,\Delta t_i\right)^2\right]$, where the time distance is measured in units of the mean synodic period of the Carrington system (27.2753 days). The advantage of this method is that it produces smoother, easier to analyze maps (data gaps in CR maps are also filled in by other maps). The choice of the kernel width (6.8 days = 1/4 of a CR) is dictated by the considerations of the intrinsic longitudinal uncertainty of the CR maps discussed above, and by the mean angular distance between the spacecraft. Indeed, the FWHM of any kernel, once the weights are normalized, has as a lower limit the difference in time before one spacecraft has the same view as another. 

Fig.~\ref{fig:comb_ex} illustrates the weighted averaging of this second method for a single longitude (45$\degr$). Since the chosen kernel width is longer than $\sim 5$ days, we would need $\sim 2$ days more of observations with respect to those needed to obtain a combined near-synchronic map with the first method (``Joined map''). Even so, as we can see from Fig~\ref{fig:comb_ex}, the CR map slice contributing mostly to the ``Merged map'' slice (black line) is the one with only 1.1 days distance from the reference time (slice of STB/E, cyan line), thus inside 5 days.  
\end{enumerate}
The maps resulting from the application of these two methods are shown in Fig.~\ref{fig:comb3}. In our analysis, we did not take into account that the two STEREO spacecraft have a non zero inclination angle ($\sim 3^\circ$) to the ecliptic plane with respect to SOHO. However, these angular differences are too small to impact the brightness of the white-light observations and/or line of sight integrations (see discussion above). 

  
\begin{figure}
  \centering
  \includegraphics[clip=true,width=\linewidth]{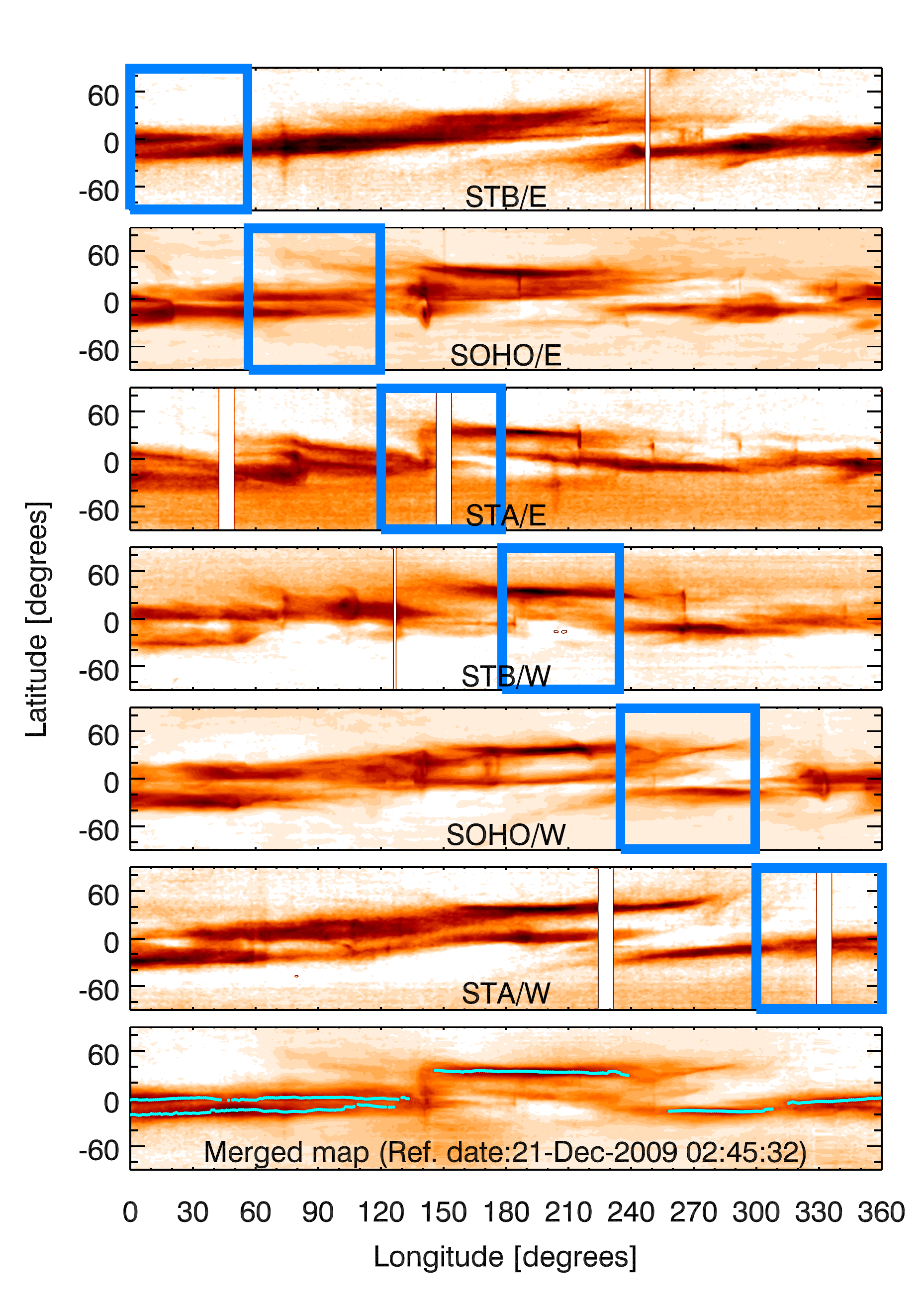}
  \caption{East and west limb Carrington maps observed by the STA/COR1, SOHO/C2, and STB/COR1 coronagraphs, for the CR 2091, at 2.5~$R_\sun$ and the ``merged'' map obtained by their combinations (bottom panel), with the position of the cusps of the streamers (cyan lines) overplotted. The blue box defines the part of the maps observed at the same time by the instruments.}
  \label{fig:obsNL1}
\end{figure}

To compare these near-synchronic maps with the solutions of the magnetic field extrapolations described in Sec.~\ref{sec:extrapolations}, we need to define which visible structures in the combined maps of Fig.~\ref{fig:comb3} are good proxies for the position of the magnetic neutral line at 2.5~$R_\sun$. As already mentioned in the introduction, it is generally accepted that streamers are the result of the line--of--sight viewing of the heliospheric plasma sheet centred at the magnetic neutral line \citep{koomen}. Hence, we compare the position of the streamers in the observed Carrington maps by taking the peaks of absolute and relative maximum values of intensity above a fixed threshold at each date, with the extrapolated neutral lines, assuming that the density enhancements corresponding to the intensity peaks track the position of the magnetic neutral line. In Fig.~\ref{fig:obsNL1} we show the Carrington maps for the East and West limb from the three different coronagraphs on STA, SOHO and STB, as well as the resulting ``merged'' combined map (bottom panel). The cyan lines in the bottom panel mark the intensity peaks. The blue contour defines the part of the maps observed at the same time by the instruments and merged in the bottom map. The positions of the main intensity peaks identified in the ``joined'' combined map are not significantly different, and thus are not shown here. 

\subsection{Data Analysis}\label{obs:analysis}

As we can see from Fig.~\ref{fig:obsNL1}, at many longitudes (i.e., between $\sim 0-140^{\circ}$), two observed streamers exist at different latitudes. Besides the rotational effects (arc--like features as in Fig.~\ref{fig:obs} at $\sim 40^{\circ}$ latitude and between $\sim 240^{\circ}-290^{\circ}$ longitude), where individual structures appear to move to higher or lower apparent latitudes as they rotate away from or toward the plane of the sky \citep{liewer}, we have to consider that some of the observed features may not correspond to a classic helmet streamer but to a pseudostreamer \citep{wang6}. Both helmet streamers and pseudostreamers contribute to the brightness of the K corona, but only the former are associated with interplanetary sector boundaries and the heliospheric current sheet. The way to distinguish between a streamer or a pseudostreamer is via coronal magnetic field extrapolations. Other characteristic features of pseudostreamers, albeit more difficult to observe, are low--lying cusps and the presence of two underlying filament channels \citep{wang6}. In the case of our data set, we do not see any of these observational signatures that could have helped distinguishing between a streamer or a pseudostreamer without resorting to extrapolations. Since we are unable to determine, observationally, if one of the two cyan lines in the bottom panel of Fig.~\ref{fig:obsNL1} represents a pseudostreamer, we will compare both lines with the extrapolated neutral lines. 

For the completeness of the discussion, we report that \citet{noci} were already observing the existence of streamers which had a bifurcated aspect in the \ion{O}{vi} image with the UltraViolet Coronagraph Spectrometer \citep[UVCS,][]{kohl} on-board SOHO, appearing to consist of two substreamers, or rather three at lower heliocentric distances ($\sim 2$~$R_\sun$). They suggested and observed in the \ion{Fe}{xiv} the existence of a quadrupolar magnetic configuration at the coronal base with three associated current sheets which give a signature in the \ion{O}{vi} observations but not in Ly-$\alpha$ (except, perhaps, at very low heights).

From now on, we refer to the enhancements observed in the Carrington maps as observed peaks of intensity, generically. In Fig.~\ref{fig:obsNL1}, we can also observe other features as the CMEs, appearing as vertical and sudden brightenings (see, for example, at $\sim 75^\circ$ or $\sim 140^\circ$).

\section{Extrapolations}\label{sec:extrapolations}

There exists a variety of magnetic field extrapolation techniques such as PFSS, Non--Linear Force--Free Field (NLFFF), magneto--frictional and full MHD approaches. A detailed description of these techniques can be found in \citet{mackay}. Here, we perform PFSS extrapolations, starting from different sets of photospheric data (described in detail later in this Section), and evaluate their performance by comparing the extrapolations with near-synchronic coronagraphic white-light CR maps of the corona near the source surface. We compare the position of the streamers in the Carrington maps with the neutral lines obtained from four different sets of calculations, named, Method1, Method2, Method3 and Method4. The characteristics of the four models are resumed in Table~\ref{table:2}.

The first two methods (Method1 and 2) use magnetic field extrapolation of synoptic magnetograms and give as result a unique source surface synoptic chart for a Carrington Rotation, i.e. we have one neutral line for each of these two methods to compare with the observations. The other two methods, Method3 and 4, instead, use synchronic (or time--evolved) photospheric maps of the magnetic field and can produce instantaneous or six hours maps of the coronal magnetic field. Hence, it is possible to have a coronal map for each instant of the Carrington Rotation. We choose to retrieve the magnetic neutral line with Method3 and Method4 at three days during the CR as explained below. 

For Method1, we use the magnetic neutral line from the WSO on--line archive (http://wso.stanford.edu/). For the other methods, we derive the neutral line via extrapolations. The four methods of extrapolation are as follows.

Method1: the method chosen uses a PFSS extrapolation from the Wilcox Solar Observatory (WSO) synoptic maps and it is described in the works of \citet{schatten,altschuler,hoeksema}. It assumes that the field in the photosphere is radial and forced to be radial at the source surface (placed at 2.5~$R_\sun$) to approximate the effect of the accelerating solar wind on the field configuration. The result of this extrapolation is a source surface synoptic chart for each Carrington Rotation. The range of dates on which the synoptic photospheric map is built is the one of the CR 2091 (7 December 2009 -- 3 January 2010). In particular, the days of contributing magnetograms are the 17th--20th and 22nd--24th of December 2009, and 3rd--4th of January 2010. The missing data are interpolated.

Method2: the method starts again with the WSO synoptic photospheric maps but uses a slightly different PFSS extrapolation following the general method and polar--field correction of \citet{wang3}. This correction enhances the polar field strength that is meant to better reproduce the open flux in the interplanetary medium, and also allows the surface field to depart from strict radiality.

Method3: we perform the same PFSS extrapolation as in Method2 but starting from the Michelson Doppler Imager (MDI) observations of the photosphere. To obtain a more realistic estimate of the global photospheric magnetic field distribution, Method3 applies a flux transport model to the photospheric data described in \citet{schrijver}. With this extrapolation method, we produce a unique map of the solar corona every six hours, and we choose three days, one at the beginning, one at the middle, and one at the end of the CR 2091 to obtain a neutral line to compare with the observations. The chosen dates are: 7 December 2009, 20 December 2009, and 3 January 2010.

Method4: we use a PFSS extrapolation as in Method2 and Method3, from the Air Force Data Assimilative Photospheric Flux Transport model \citep[ADAPT,][]{adapt}. The ADAPT model generates more realistic global solar photospheric magnetic field maps, starting from NSO/GONG data, in our case. ADAPT produces ensemble synchronic predictions (i.e., multiple maps) and, in particular, for our work, it produces twelve solutions (depending on different choices of the physical parameters in the simulation, to cover the uncertainty related to those parameters), for each predicted day of the Carrington map. We choose to have extrapolated data for the same three days as for Method3. 

\begin{table*}
\caption{Characteristic of the four extrapolation methods used in this work. References: $^1$\citet{schatten}, $^2$\citet{altschuler}, $^3$\citet{hoeksema}, $^4$\citet{wang3}, $^5$\citet{schrijver}, $^6$\citet{adapt}.}
\label{table:2}      
\centering
\begin{tabular}{c c c c c}
\hline\hline
            & Method1 & Method2 & Method3 & Method4 \\
\hline\hline       
\vspace{1mm}
Photospheric magnetic maps & WSO & WSO & SOHO/MDI & NSO/GONG \\
\vspace{1mm}
Synoptic or synchronic & Synoptic & Synoptic & Synchronic & Synchronic \\
\vspace{1mm}
Treatment of photospheric maps & None & None & Flux transport model$^{5}$ & ADAPT flux transport model$^{6}$ \\
References for PFSS Extrapolation Method & 1,2,3  & 4 & 4 & 4 \\
\hline
\end{tabular}
\end{table*}

\section{Results and Discussion}\label{sec:results}
 
\begin{figure}
\centering
\includegraphics[clip=true,angle=180,width=\linewidth]{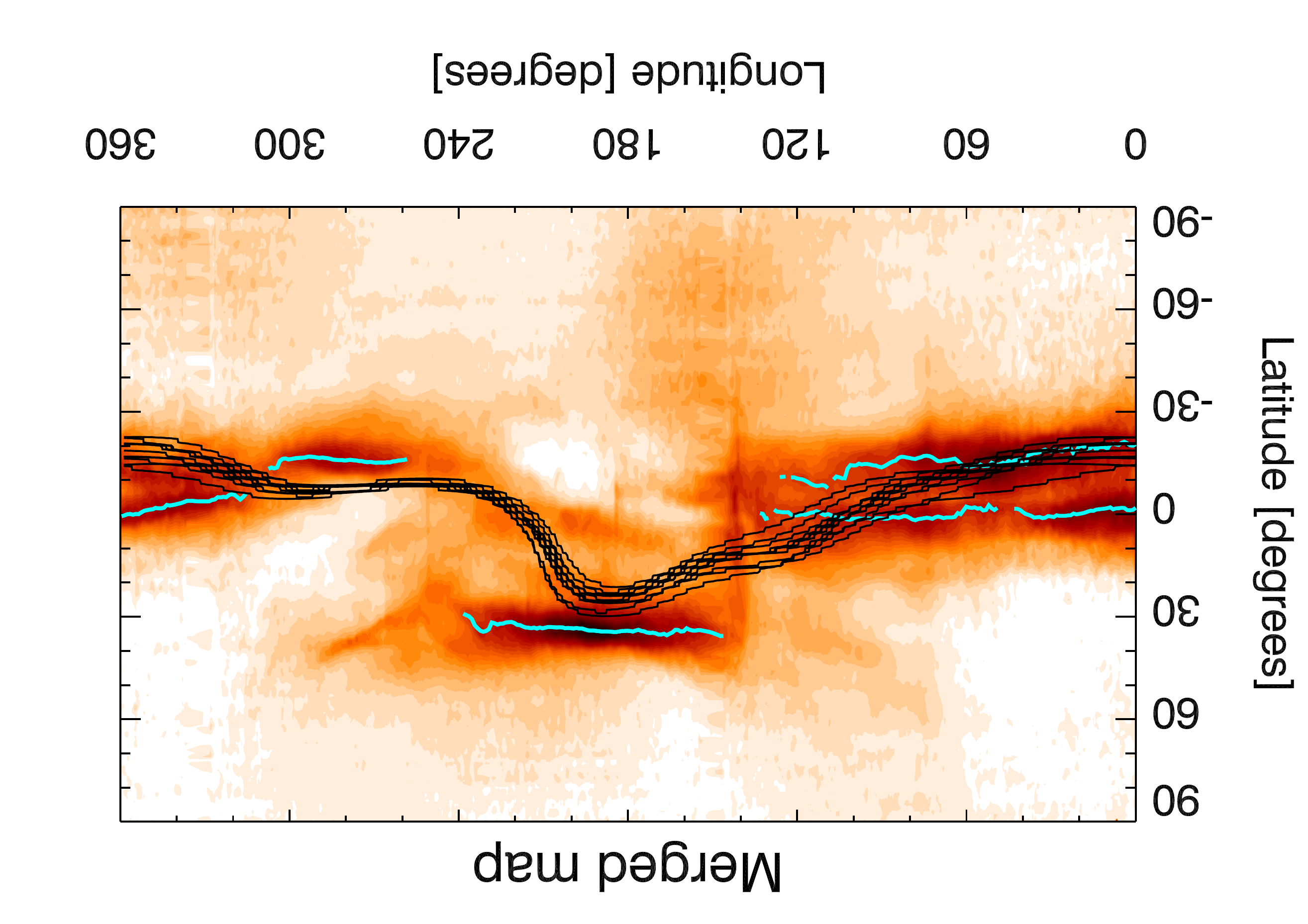}
\caption{Merged Carrington map with the streamer axes (cyan lines) and calculated neutral lines (black lines) from Method4 for 20 December 2009.}
\label{fig:adapt}
\end{figure}

We start with the comparison between the streamer intensity peaks and the neutral lines calculated with Method4. As we explained in Sec.~\ref{sec:extrapolations}, this method produces twelve solutions for the coronal magnetic field and, hence, twelve neutral lines, for each day chosen to perform an extrapolation. In Fig.~\ref{fig:adapt} we show the neutral lines (black lines) obtained for the 20 December 2009 extrapolation. We plot the peaks of streamer intensity (cyan lines) on the merged Carrington map from the bottom panel of Fig.~\ref{fig:obsNL1}. The differences among the twelve extrapolated neutral lines are no more than $20\degr$. Although we show the results for only one of the three days of extrapolations, the same finding holds for the other two days. Therefore, we show only one of these extrapolations from now on. To measure the difference between streamers and calculated HCS, we use the absolute value of the latitudinal difference, $|Lat_{obs}-Lat_{ext}|$, between the two features at each longitude. Since there exist two streamers belts between $\sim 0-140^{\circ}$, we perform the calculation of these differences for two cases, one considering the points with the highest latitudes $(Lat_{obs1}$,Case 1), and the other considering the points with the lowest ones $(Lat_{obs2}$, Case 2). For Method4, we then plot the extrapolated neutral line that has the best fit with one of the two lines representing the observations, calculated as the mean value of $|Lat_{obs}-Lat_{ext}|$.

\begin{figure}
\centering
\includegraphics[clip=true,angle=180,width=\linewidth]{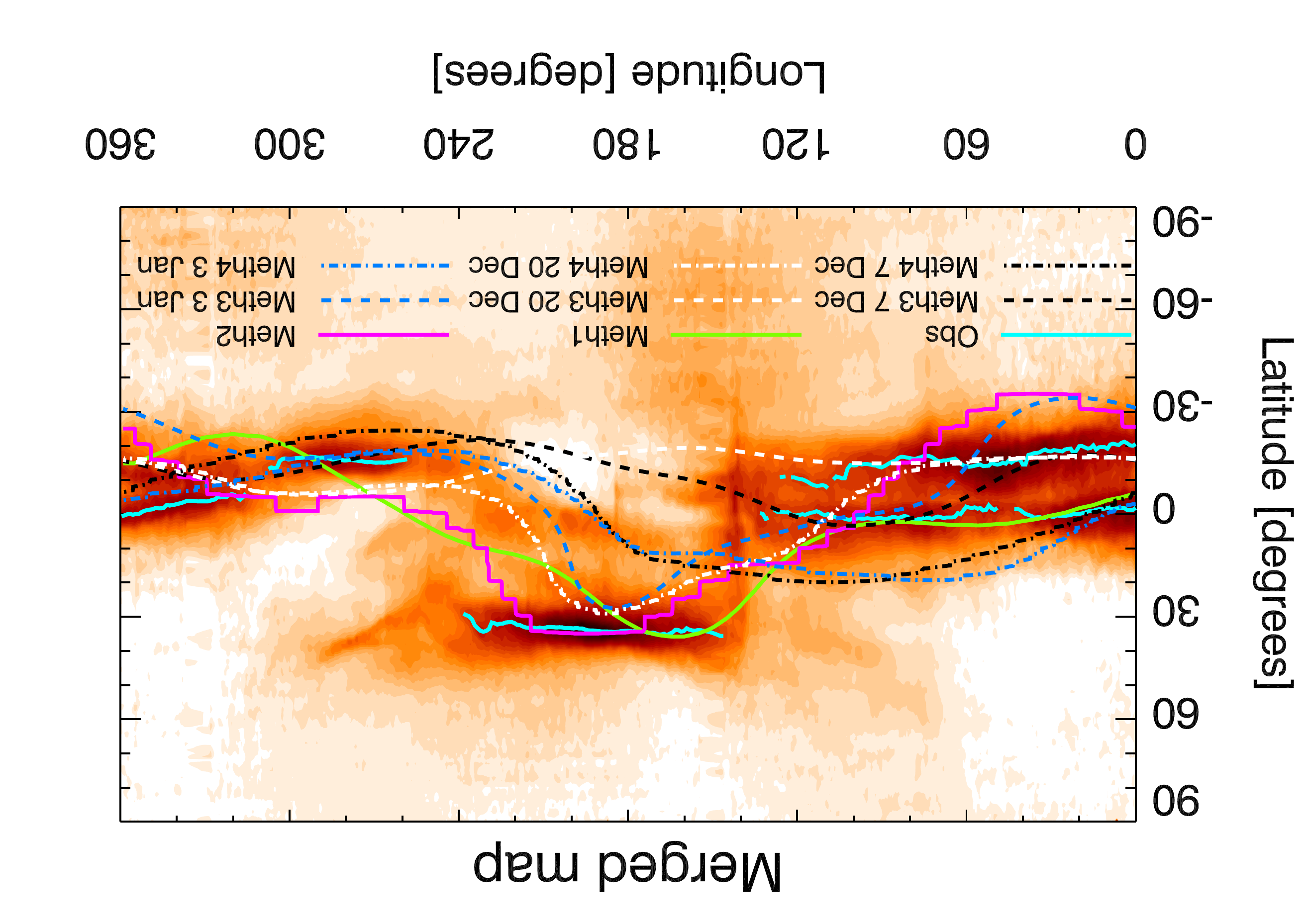}
\caption{Merged Carrington map with overplotted the observed peaks of intensity (cyan lines) and the neutral lines obtained from the four methods of extrapolation described in Sec.~\ref{sec:extrapolations}. The colour coding is given in the figure legend.}
\label{fig:results}
\end{figure}

In Fig.~\ref{fig:results} we compare the streamer axes with the neutral lines derived from all four extrapolation methods on the merged Carrington map. For Method1 and Method2 we plot one neutral line (solid green and magenta lines, respectively), while for Method3 and Method4 we plot 3 neutral lines corresponding to the 3 days chosen to perform the extrapolations (three dashed lines for Method3 and three dashed-dotted lines for Method4, plotted with different colours to distinguish among the days: black for 2009 December 7, white for the 2009 December 20, and blue for the 2010 January 3). 

Even though the extrapolations use the same photospheric magnetic field data (WSO synoptic maps) and the PFSS methods are very similar, Method1 and Method2 provide different results. The main difference lies between $0-100^{\circ}$ longitudes, where we also have two streamer axes, one at $\sim 0^{\circ}$ latitude and the other at $\sim -20^{\circ}$: the neutral line from Method1 overlaps the observation at $\sim 0^{\circ}$ latitude, while the Method2 neutral line overlaps the one at $\sim -20^{\circ}$ latitude.  

Regarding Method3 and Method4, we note that the neutral lines obtained from the extrapolations computed on three different days can be very different. This is because the magnetic field in the photosphere can change during a solar rotation. Indeed, there a strong bipolar region emerges between $(60^{\circ}, -60^{\circ})$ Carrington coordinates at the end of the Carrington rotation. 

\begin{figure}
\centering
\includegraphics[clip=true,angle=180,width=\linewidth]{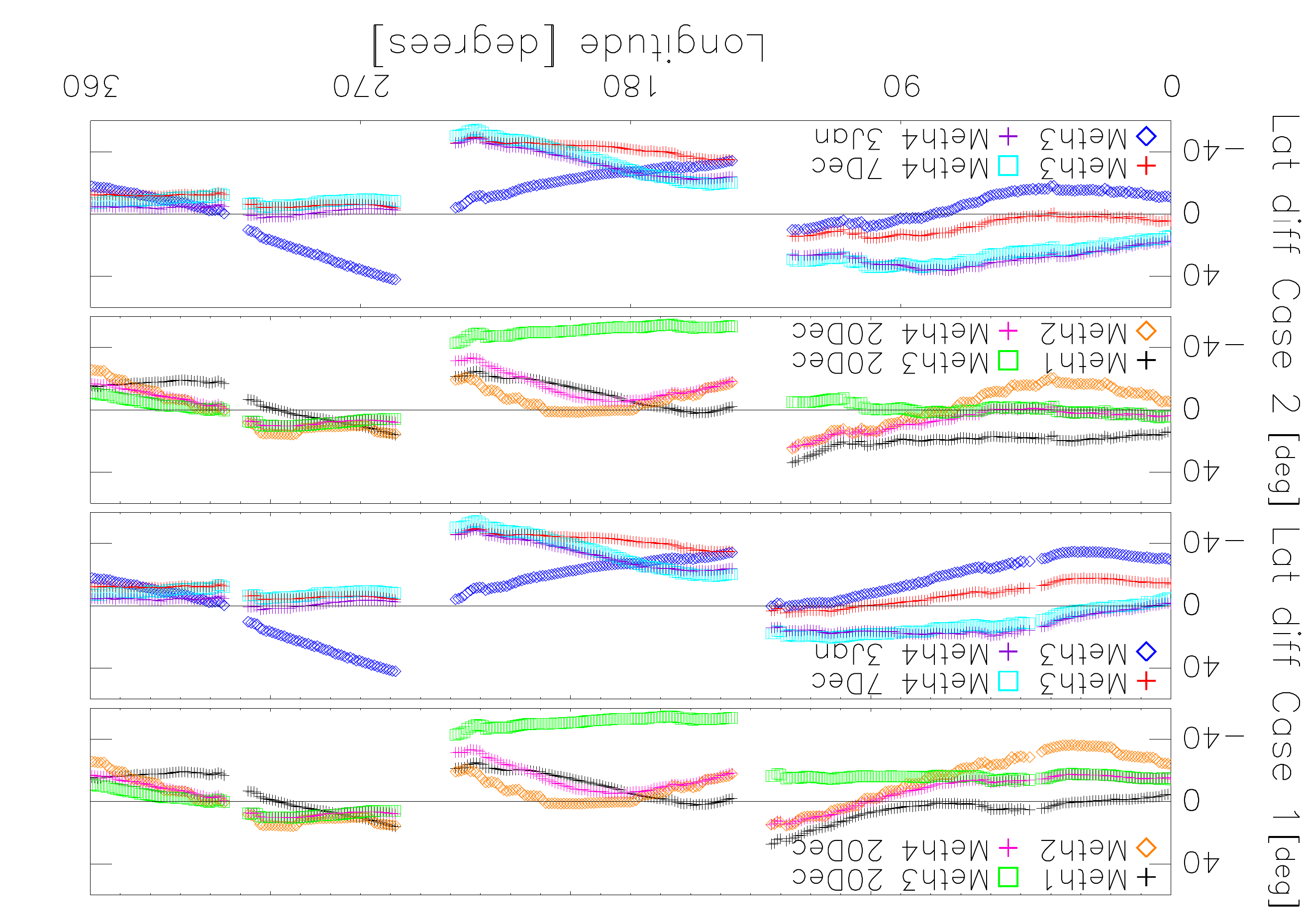}
\caption{Latitudinal differences between the extrapolated neutral lines and the two streamer axes, as function of the longitude. The colour coding is given in the figure legend.}
\label{fig:diff}
\end{figure}

To evaluate the longitudinal dependence of the accuracy of the extrapolations, we plot in Fig.~\ref{fig:diff} the quantity $Lat_{obs}-Lat_{ext}$ as function of the longitude for the two streamer axes, $Lat_{obs1}$ (top panels) and $Lat_{obs2}$ (bottom panels). Table~\ref{table:1} summarizes the means of the absolute value of the differences, and the standard deviations of these differences. We note that the two metrics are very well correlated.

We see that four neutral lines (Method1 for Case 1, Method2 for Case 2, and Method4 (20 Dec) for both Case 1 and 2) give a good approximation of the streamers position with a mean error of $\sim 9-12^\circ$ and a standard deviation of $\sim 11-12^\circ$, with respect to the other methods. Among these four, Method1 is the method that gives the neutral line with the absolute best agreement with the observations, assuming that the streamer between $0-120^\circ$ longitude has a latitude of $\sim 0^\circ$. If we compare the values of the means and the standard deviations in Table~\ref{table:1} for Case 1 and Case 2, for all the methods, there is no evidence of a better agreement of the extrapolated neutral lines with one or the other streamer we observe between $0-120^\circ$ longitude. Thus, neither the observations nor the extrapolations allow us to discern the presence of a pseudostreamer in this CR. 
 
We find a good agreement with the neutral lines extrapolated from Method1 and Method2 (based on synoptic maps) but also from Method4 (based on synchronic map, for 20 December 2009). If we look in detail at the photospheric synoptic maps used in the first two methods we see that the days of magnetogram observations contributing to the synoptic map are the 17th--20th and 22nd--24th of December 2009, and 3rd--4th of January 2010, with the missing data interpolated. Hence, we have a synoptic map of the photospheric magnetic field build using mainly data acquired around the 20th of December that is the date we have chosen as the reference time to build the near-synchronic coronal map.
The major uncertainties in the comparison of the neutral lines from both Method1 and Method2 and the observed streamers are at small longitudes ($30-90^\circ$, which correspond to the dates of 3--4 January 2010) where the photospheric magnetic field changes due to the emergence of a strong bipolar region. 

\begin{table*}
\caption{The means of the absolute value of the differences between the extrapolated neutral lines latitudes and that of the observed streamers, and the standard deviation of these differences, for the two streamer axes. Discussion is in the text.}
\label{table:1}      
\centering
\begin{tabular}{c c c c c c c c c c}
\hline\hline
& & Meth1 & Meth2 &         & Meth3    &         &         & Meth4    &        \\
& $(^\circ)$ &       &       & (7 Dec) & (20 Dec) & (3 Jan) & (7 Dec) & (20 Dec) & (3 Jan) \\    
\hline\hline  
Case 1 & Mean  & 9 & 14 & 18 & 23 & 20 & 18 & 12 & 16 \\
& Std   & 12 & 15 & 17 & 21 & 20 & 21 & 11 & 20 \\
\hline
Case 2 & Mean  & 15 & 10 & 18 & 18 & 16 & 23 & 10 & 22 \\
& Std   & 16 & 12 & 21 & 24 & 18 & 27 & 12 & 26 \\
\hline
\end{tabular}
\end{table*}

\section{Conclusions}\label{sec:conclusions}

The aim of this study is to find a fast and reliable method to validate systematically the various method of coronal magnetic field extrapolations. For this purpose, we determine the position of the coronal streamers by their intensity peaks and compare them with the location of the magnetic neutral lines obtained from four different coronal magnetic field extrapolation methods. The comparison is based on the assumption that the intensity enhancements track the position of the streamers and the associated magnetic neutral line at $2.5$~R$_\sun$. This assumption is not always true, since a denser sheet of plasma visible as a bright enhancements in the white--light observations, can be also related to a pseudostreamer that does not enclose a current sheet. Observationally, at least for CR2091, we can not distinguish between streamers and pseudostreamers but we can still derive useful information on the reliability of the extrapolations.

The improvement of our study with respect to the previous attempts of comparing the position of the streamers in white--light Carrington maps and extrapolated neutral lines (see Sect.~\ref{intro}) is in the combination of SOHO/LASCO--C2, STA and STB/SECCHI--COR1 Carrington maps at $2.5$~R$_\sun$ for Carrington Rotation CR 2091 to obtain a synoptic map of the solar corona with the minimum amount of temporal evolution and then compare the coronal structures visible in this coronagraphic map with magnetic field extrapolations.

The four extrapolation methods are described in Sec.~\ref{sec:extrapolations}. The first two methods are based on synoptic photospheric maps and provide one neutral line to be compared with the observations. Method3 and Method4, instead, produce instantaneous coronal magnetic field maps starting from synchronic photospheric maps for each instant of the CR. For these two methods, we choose the extrapolated magnetic neutral line for three days during the CR. Moreover, Method4 produces twelve neutral lines for each day. We find that the differences among these neutral lines are too small to let us distinguish among them through a comparison with the observations. We do not need such a high resolution in the extrapolations for this kind of analysis.

Considering the results of Method3 and Method4, we find that the neutral lines obtained from three different days during the Carrington rotation can be very different. This is because the photospheric field data may change on a time scale shorter than thirteen days (minimum temporal distance between the dates chosen). This result underlines the importance of reducing the time needed to scan the corona by, for example, combining images from instruments looking at the Sun from different viewing angles. After Solar Orbiter is launched, there will be several coronagraphs from space, i.e. Metis \citep{antonucci} on Solar Orbiter itself, ASPIICS on PROBA--3 \citep{lamy,renotte} to coordinate for joint observation campaigns. We have to take into account that for some instruments, like Metis that will observe also out of the ecliptic plane, there will be the need of other techniques, like tomography, to compare the data with the other coronagraphs. Tomography is of course a valid approach also from inside the ecliptic but its use is beyond the scope of this work. On board Solar Orbiter, there will also be a magnetograph, the Polarimetric and Helioseismic Imager \citep[PHI,][]{phi1,phi2} providing magnetograms of the solar photosphere. In this way, we will also have simultaneous maps of the photospheric magnetic field  available over more than just the solar surface visible from Earth.
 
Comparing the neutral lines resulting from the four methods and the positions of the streamers, we find a good agreement for Method1, Method2, and Method4 (performed on 20 Dec 2009). We notice that all three methods start from photospheric maps of the magnetic field (synoptic or synchronic) built with data acquired around the same day we have chosen as the reference time to build the merged coronal map from the contributions of the Carrington maps of the three spacecraft. This is not a necessary condition, however, to have a good extrapolation, since the magnetic neutral line extrapolated using Method3 (on 20 December 2009) gives a bigger error, compared with the observations (see Table~\ref{table:1}), at least for CR2091. A comparison in details of the different extrapolation methods will be the subject of a subsequent paper.

The method described in this paper has clear advantages in its simplicity and in the availability of the observations and extrapolations but it can fail in some situations like when there are multiple white-light features at different latitudes, creating an ambiguity on their interpretation. Indeed, in our analysis we get a decent overall agreement at the Carrington longitudes for which the position of the streamers is unique and a significant disagreement at Carrington longitudes where multiple white-light features arise. The current set of extrapolations reveal to be not useful beyond a top-level comparison with the coronagraphic images (i.e., the existence of a streamer at a given position angle). They seem to lack robust information on the field to go beyond that. Also, this study shows that different magnetograms sources can also disagree.  

We plan to re-do the multi-viewpoint analysis presented in this paper with other CR maps to cover, at least, one solar cycle. It would be interesting also to compare the combined quasi-synchronic maps with results from other extrapolations than the ones we used for this work. We also plan to identify in the white-light maps further visible structures (for example, the position of the coronal holes) to compare them with the results of the extrapolations.

\begin{acknowledgements}

Wilcox Solar Observatory data used in this study was obtained via the web site http://wso.stanford.edu at 2016:12:09\_03:10:30 PST courtesy of J.T. Hoeksema. The Wilcox Solar Observatory is currently supported by NASA. 
The SOHO/LASCO data used here are produced by a consortium of the Naval Research Laboratory (USA), Max-Planck-Institut for Sonnensystemforschung (Germany), Laboratoire d'Astrophysique Marseille (France), and the University of Birmingham (UK). SOHO is a project of international cooperation between ESA and NASA.
C.S., V.A., A.B., S.D., D.S., R.S., E.A., V.D.D., S.F., F.F., F.L., G.Na., G.Ni., P.N., M.P., M.R., D.T. and R.V. acknowledge the support of the Italian Space Agency (ASI) to this work through contract ASI/INAF No. I/013/12/0. A.V. is supported by NRL N00173-16-1-G029 and NASA NNX16AH70G grants.

\end{acknowledgements}

\bibliographystyle{aa}

\end{document}